\documentclass[fleqn,usenatbib]{mnras}
\pdfoutput=1
\usepackage{graphicx}
\usepackage{color}
\usepackage{amsmath}
\usepackage{amssymb}
\usepackage{bm}
\usepackage{siunitx}	
\usepackage{newtxtext,newtxmath}
\usepackage[T1]{fontenc}
\usepackage{ae,aecompl}
\newcommand{{\msun}}{$\,\mathrm{M}_{\odot}$}
\DeclareSIUnit\mass{M}
\DeclareSIQualifier\solar{\odot}
\newcommand{{\simu}}{\emph{\,uniform}}
\newcommand{{\simp}}{\emph{\,partially turbulent}}
\newcommand{{\simt}}{\emph{\,fully turbulent}}
\usepackage{hyperref}
\usepackage{etoolbox}
\makeatletter
\patchcmd\@combinedblfloats{\box\@outputbox}{\unvbox\@outputbox}{}{%
   \errmessage{\noexpand\@combinedblfloats could not be patched}%
}%
 \makeatother

\title{The role of initial magnetic field structure in the launching of protostellar jets}
\author[Gerrard et al.]{
Isabella A. Gerrard$^{1,2}$, Christoph Federrath$^{1}$ \& Rajika Kuruwita$^{1}$
\\
$^{1}$Research School of Astronomy and Astrophysics, Australian National University, Canberra, ACT 2611, Australia\\
$^{2}$Monash Centre for Astrophysics, Monash University, Clayton, VIC 3800, Australia\\
}
\date{Accepted XXX. Received YYY; in original form ZZZ}
\pubyear{2019}
\begin{document}
\label{firstpage}
\pagerange{\pageref{firstpage}--\pageref{lastpage}}
\maketitle

%%%%%%%%%%%%%%%%%%%%%%%%%%%%%%%%%%%%%%%%%%%%%%%%%%

\begin{abstract}
Magnetic fields are known to play a crucial role in the star formation process, particularly in the formation of jets and outflows from protostellar discs. The magnetic field structure in star forming regions is not always uniform and ordered, often containing regions of magnetic turbulence. We present grid-based, magneto-hydrodynamical simulations of the collapse of a 1{\msun} cloud core, to investigate the influence of complex magnetic field structures on outflow formation, morphology and efficiency. We compare three cases: a uniform field, a partially turbulent field and a fully turbulent field, with the same magnetic energy in all three cases. We find that collimated jets are produced in the uniform-field case, driven by a magneto-centrifugal mechanism. Outflows also form in the partially turbulent case, although weaker and less collimated, with an asymmetric morphology. The outflows launched from the partially turbulent case carry the same amount of mass as the uniform-field case but at lower speeds, having only have 71\% of the momentum of the uniform-field case. In the case of a fully turbulent field, we find no significant outflows at all. Moreover, the turbulent magnetic field initially reduces the accretion rate and later induces fragmentation of the disc, forming multiple protostars. We conclude that a uniform poloidal component of the magnetic field is necessary for the driving of jets.	
\end{abstract}


\section{Introduction} \label{intro}
Stars form in giant molecular clouds (GMCs) \citep{mac-low04, elmegreen04, mckee07, hennebelle12}, which are ubiquitously permeated by magnetic fields \citep{crutcher12} and are observed to be inherently turbulent \citep{larson81,dubinski95,ossenkopf02,heyer04,roman-duval11,hennebelle12}. In the last decade it has become clear that turbulence and magnetic fields are vital prerequisites for star formation \citep{mac-low04,mckee07,federrath12,joos13,li14,padoan14,federrath15}. 
Observations of large and small scale magnetic field structures in GMCs are inferred by mapping dust polarisation. Dust grains align themselves perpendicular to the local magnetic field lines \citep{lazarian07}, and the resultant polarised thermal emission can then be observed as a tracer of the field structure \citep{planck-collaboration16}. Observations of dust polarisation in GMCs have recently revealed turbulent magnetic field structures in star forming cores \citep{girart13, ching17,hull17, king18}, as well as cores threaded with distinctly uniform fields \citep{chapman13, pattle17}. Although the intricacies of the dust alignment mechanism make extracting the \emph{exact} magnetic field morphology from polarisation maps difficult \citep{king18}, the technique has demonstrated that magnetic fields in star forming cores and throughout the universe are complex structures that vary from uniform and ordered to tangled and chaotic \citep{cox18}.

The formation of jets and outflows during star formation has been proposed as a partial solution to some of the ongoing problems of star formation theory \citep{pudritz86,crutcher91}. Bipolar jets and outflows carry angular momentum away from the protostellar disc, which accounts for the relatively slow rotation rates of stars \citep{pudritz07,joos12,frank14}. These jets also transport mass away from the accretion disc, which impacts the \emph{star formation efficiency} (SFE) and may help to explain low star formation rates \citep{price09,krumholz14,padoan14, federrath15}. Driven by magnetic pressure \citep{lynden-bell03}, and in some cases accelerated by the magneto-centrifugal mechanism \citep{blandford82}, these bipolar outflows and jets can extend up to several parsecs \citep{stojimirovic06}. Herbig-Haro objects with distinctive bipolar outflows are observational confirmation of this phase of star formation \citep{bally01,wu04}. The distinction between jets and outflows in the literature is somewhat ambiguous; while it is true that all jets are outflows of a particular kind, not all outflows are fast, collimated and \emph{jet-like}. We will refer to high speed outflows exhibiting high degrees of collimation as \emph{jets} and those with less distinct structure as \emph{outflows}, although this is a purely qualitative distinction.

The majority of the computational work simulating the launching of jets and outflows during star formation has used ordered magnetic fields of varying strength \citep{banerjee06,shang06,machida06,price07,machida08,hennebelle08,price12b,bate14}. We use the term \emph{ordered} to refer to fields that are uniform and aligned with the rotation axis of the accretion disc. There has been some work done on the effects of a \emph{misaligned} magnetic field (one which is still uniform but the angle between the disc rotation axis and the mean component of the field is non-zero) \citep{hennebelle09,joos12,krumholz13,liZY13,lewis15, lewis17}. \citet{lewis15} used smoothed particle magneto-hydrodynamics to model the collapse of a magnetised cloud core and found that collimated jets are produced with shallow alignment angles ($<10^{\circ}$), and that no discernible outflows were produced with angles greater than $60^{\circ}$. Their work is in agreement with that of \citet{ciardi10} who carried out adaptive mesh refinement (AMR), magneto-hydrodynamical (MHD) simulations with similar results, showing that all outflowing material was suppressed if the field was aligned perpendicular to the rotation axis. The orientation of the field relative to the rotation axis also effects the formation and survival of protostellar discs \citep{liZY14, gray18}. However, the role of \emph{turbulent} magnetic field structures has not been quantified in previous work.

Here we carry out a controlled numerical experiment to ascertain what effect an initially turbulent magnetic field has on the early stages of jet and outflow formation. This experiment has been designed such that the only parameter that varies is the \emph{structure} of the magnetic field. Our numerical methods are described in Section \ref{meth}, the results and analysis of which are presented in Section \ref{res}. The implications of these results are discussed in Section \ref{dis}. In Section \ref{conc} we present our conclusions and summarise the outlook for future work.
\section{Methods} \label{meth}
We carry out MHD simulations of protostellar disc and star formation. We compare a set of three different simulations, which only differ in their initial magnetic field structure. The first is a control simulation in which the magnetic field is initially ordered and aligned with the z-component of angular momentum, hereafter {\simu}. The second is constructed so as to affect equally ordered and unordered components of the magnetic field (see Section \ref{partial}) with the mean field component aligned as in the ordered case, hereafter {\simp}. The third simulation starts with a fully turbulent field without any ordered, mean component, hereafter {\simt}.

We compare these three simulations to study the effects of the turbulent field structure. All other aspects of the setup are identical in the three simulations, closely following those of \citet{federrath14} and \citet{kuruwita17}, which are summarised below.

\subsection{FLASH}\label{flash}
We use FLASH, a grid-based, adaptive mesh refinement (AMR) code \citep{fryxell00, dubey08}. The refinement criteria is based on the Jeans length, such that the mesh is refined in regions where the Jeans length is resolved within less than 32 grid cells \citep{sur10, federrath11a, federrath14}. This ensures that the Jeans length is always resolved everywhere in at least 32 grid cells. We use the positive-definite HLL3R Riemann scheme \citep{waagan11} to solve the standard set of three-dimensional, ideal MHD equations:
\begin{align}
	& \frac{\partial \rho}{\partial t} + \nabla \cdot \left(\rho \ \bm{v}\right) = 0,\\
	\nonumber \\
	& \rho \left(\frac{\partial}{\partial t} + \bm{v} \cdot \nabla \right) \bm{v} = \frac{\left(\bm{B} \cdot \nabla \right)\bm{B}}{4\pi} - \nabla P_{\mathrm{tot}} + \rho \ \bm{g}, \\
	\nonumber \\
	& \frac{\partial E}{\partial t} + \nabla \cdot \left(\left(E+P_{\mathrm{tot}}\right)\bm{v} - \frac{\left(\bm{B} \cdot \bm{v} \right)\bm{B}}{4\pi}\right) =\rho \ \bm{v} \cdot \bm{g},\\
	\nonumber \\
	& \frac{\partial \bm{B}}{\partial t} = \nabla \times \left(\bm{v} \times \bm{B} \right),\\
	\nonumber \\
	& \nabla \cdot \bm{B} = 0\ ,
\end{align}
where $\rho$, $\bm{v}$, $\bm{B}$, $\bm{g}$ and $E$ denote the gas density, velocity, magnetic field strength, gravitational acceleration of the gas and total energy density (internal, plus magnetic, plus kinetic), respectively. 

The div(B) constraint is enforced by a parabolic divergence cleaning method following the basic concept of \citet{marder87}. However, we have also tested divergence cleaning schemes that combine parabolic and hyperbolic cleaning, by \citet{dedner02} and with recent modifications by \citet{tricco16}. For our particular problem setup, these schemes provide similar results and performance. The div(B) cleaning scheme used here was tested in \citet{waagan11}  and yields reasonable results; i.e., while none of these cleaning schemes yield div(B) to machine precision everywhere and at all times (as opposed to Constrained Transport; see \citet{evans88} and \citet{gardiner05}), the errors introduced due to div(B) waves are negligible and do not affect our results and conclusions.

$P_{\mathrm{tot}}$ denotes the sum of the magnetic and thermal pressure, where the magnetic pressure is given by
\begin{align}
	P_{\mathrm{B}} = \frac{|\bm{B}|^2}{8 \pi} \ .
\end{align}
To define the thermal pressure $P_{\mathrm{th}}$, we adopt a piecewise polytropic equation of state with $\Gamma$ values derived from the detailed radiation-hydrodynamic simulations of \citet{masunaga00}:
\begin{align}
	P_{\mathrm{th}} = K \rho^{\Gamma}\ ,
\end{align}
where $\Gamma$ is defined by
\begin{align}
	\Gamma = 
	\begin{cases}
		&1.0\text{ for } \ \ \ \ \ \ \ \rho \le \rho_1\equiv \SI{2.50e-16}{\gram\per\centi\metre\cubed},\\
		&1.1\text{ for } \rho_1< \rho \le \rho_2 \equiv \SI{3.84e-13}{\gram\per\centi\metre\cubed},\\
		&1.4\text{ for } \rho_2< \rho \le \rho_3 \equiv \SI{3.84e-8}{\gram\per\centi\metre\cubed},\\
		&1.1\text{ for } \rho_3< \rho \le \rho_4 \equiv \SI{3.84e-3}{\gram\per\centi\metre\cubed},\\
		&5/3\text{ for } \ \ \ \ \ \ \rho>\rho_4.
	\end{cases}
	\label{eos}
\end{align}
These polytropic exponents affect radiative transfer on the local cell scale, tracking the initial isothermal contraction, adiabatic heating of the first core, H$_2$ dissociation in the second collapse leading to the second core, and then return to adiabatic heating. 
The polytropic constant $K$ is then adjusted according to the density in each regime of Equation \ref{eos} such that $ K= c_{\mathrm{s}}^2$, so that the sound speed $c_{\mathrm{s}}$ and temperature are continuous functions of density. In the isothermal regime when $\Gamma = 1$, we define the sound speed in the gas to be \SI{2.0e4}{\centi\metre\per\second}, which corresponds to a temperature of \SI{11}{\kelvin} and a mean molecular weight of $2.3m_{\mathrm{H}}$. The polytropic constant is $K \equiv  \SI{4.0e8}{\square\centi\metre\per\square\second}$. For a dense molecular gas of solar metallicity, $\rho \le  \SI{2.50e-16}{\gram\per\centi\metre\cubed}$ is a good approximation for the initial isothermal evolution for a wide range of densities \citep{wolfire95,omukai05,pavlovski06,glover07a,glover07b,glover10,hill11,hennemann12,glover12}.
The gravitational acceleration of the gas, $\bm{g}$, is comprised of the gravitational potential of the gas and the contribution of any sink particles (described in Section \ref{sinks}) which may be formed. It is therefore defined as
\begin{align}
	\bm{g} = - \nabla \Phi_{\mathrm{gas}} + \bm{g}_{\mathrm{sinks}}.
\end{align}
We use a tree-based OctTree algorithm to calculate the gravitational interaction of the gas \citep{wunsch18}, and direct N-body summations to calculate the gas-sink interactions \citep{federrath14}.

\subsection{Sink Particles} \label{sinks}
\citet{truelove97} found that the Jeans length must be resolved within four grid cells to avoid artificial fragmentation. Using sink particles ensures that resultant protostars are not numerical artefacts, allows us to advance our simulations beyond the point of first collapse, and also enables us to quantify the star formation rate and SFE \citep{federrath10b}. We therefore define the density threshold for sink particle creation as
\begin{align}
	\rho_{\mathrm{sink}}=\frac{\pi c_{\mathrm{s}}^2}{4 G r_{\mathrm{sink}}^2} \ ,
	\label{rhosink}
\end{align}
where $r_{\mathrm{sink}} = 2.5 \Delta x$, $\Delta x$ being defined as the cell length on the maximum AMR level. In our simulation $\Delta x = 3.05\,$AU (see Section \ref{numres}). The critical density for sink formation at the highest level of refinement is $\rho_{\mathrm{sink}} = \SI{9.15e-11}{\gram\per\cubic\centi\metre}$, which is iteratively calculated via Equation \ref{rhosink} and  Equation \ref{eos}, as the sound speed is a function of density. Once a cell has reached this density threshold, a control volume of radius $ r_{\mathrm{sink}}$ is created around that cell. If the gas inside the control volume meets the following criteria, described by \citet{federrath10b}, a sink particle is created:
\small
\begin{enumerate}
	\item The cell with $\rho \ge \rho_{\mathrm{sink}}$ is on the highest level of AMR,
	\item No other sink particle is within $r_{\mathrm{sink}}$ of the control volume,
	\item The gas is converging from all directions, $(v_{\mathrm{r}} <0)$,
	\item The gas is bound such that $|E_{\mathrm{grav}}|> E_{\mathrm{th}} + E_{\mathrm{kin}} + E_{\mathrm{mag}}$,
	\item The volume has a gravitational minimum at its centre,
	\item The gas is Jeans-unstable.
\end{enumerate}
\normalsize 
Once a sink particle is created, it is then able to accrete gas, provided that gas has exceeded the density threshold within the accretion radius of the sink and the gas is bound and collapsing towards the sink particle. The accreted gas is removed from the MHD system and added to the sink particle, such that the total mass, the centre of mass, and the linear and angular momenta are conserved. This process is described in detail by \citet{federrath14}.

\subsection{Initial Conditions} \label{init}
\begin{figure*}
	\centerline{\includegraphics[width=\linewidth]{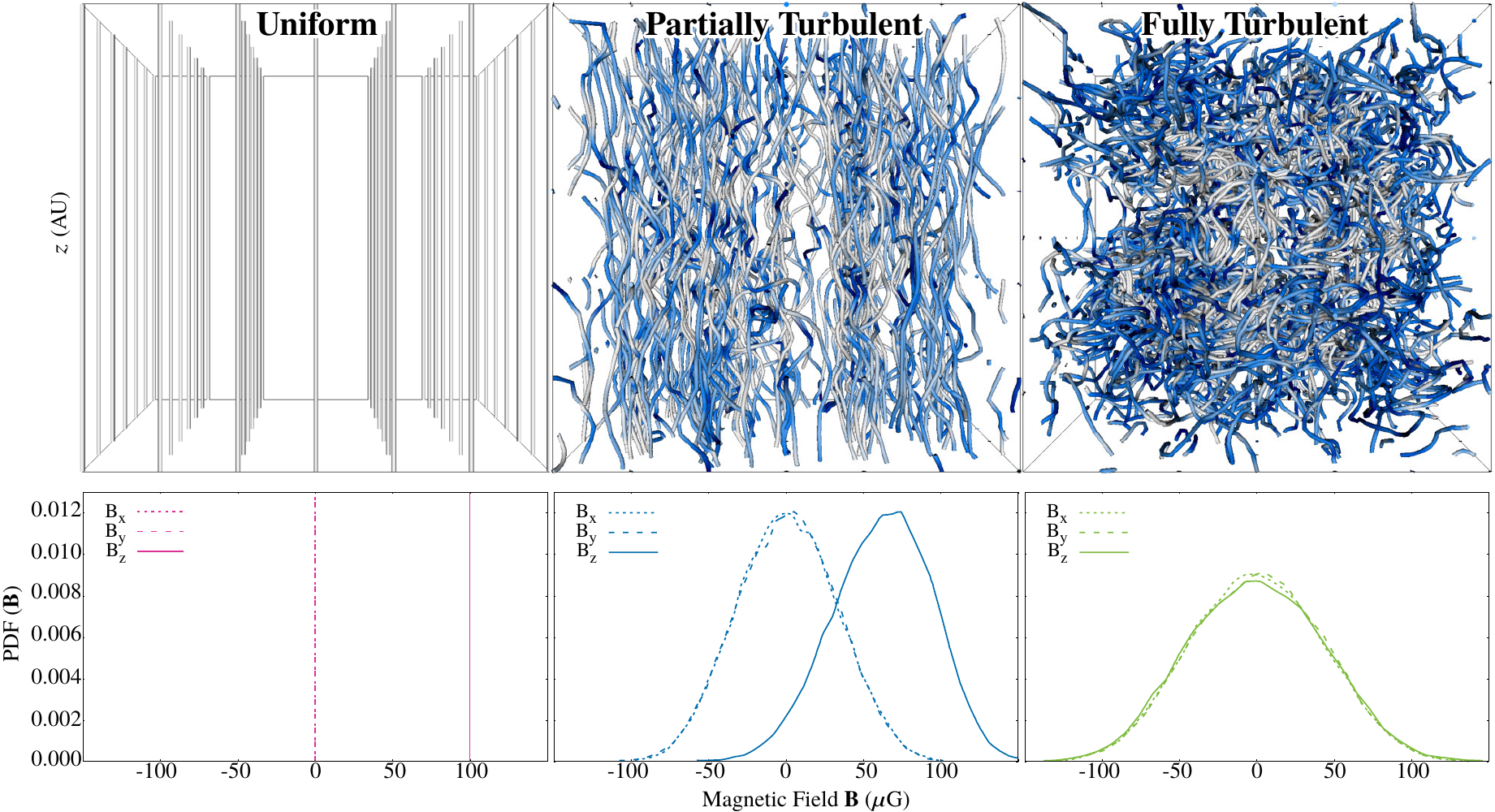}} 
  \caption{\emph{Top}: 3D visual representation of the initial magnetic field structure in our three simulation cases: Uniform (left), Partially Turbulent (middle), and Fully Turbulent (right). Each pane shows the entire simulation domain ($8000\,$AU on each side), and a sample of the field structure. The colouring of the field lines is an arbitrary consequence of the 3D rendering used and does not reflect field strength. \emph{Bottom}: Probability density functions (PDFs) of the initial magnetic field components (in $B_x$, $B_y$, $B_z$) for each of the three simulation models, respectively. The vertical lines in the {\simu} model are delta functions, i.e., $B_x=B_y=0$ and $B_z=100\,\mu\mathrm{G}$. For the {\simp} and {\simt} models, the field distributions are Gaussian as imposed by the turbulent field construction method (see Section \ref{turb} for details). Note that the total RMS magnetic field strength is identical with $100\,\mu\mathrm{G}$ in all three simulation cases, such that our study probes the consequences of different magnetic field structure (with the overall strength of the field held constant).}
  \label{initcond}
 \end{figure*}

Following initial conditions similar to those of \citet{federrath14} and \citet{kuruwita17}, our computational domain is a box of size \SI{1.2e17}{\centi\metre}, or $8000\,$AU on each side. It contains a 1{\msun} gas cloud of radius $3300\,$AU with uniform density of $\rho_0=\SI{3.82e-18}{\gram\per\cubic\centi\metre}$. So that the cloud is well defined, the gas density outside the cloud is $\rho_0/100$, with internal energy such that the cloud is in pressure equilibrium with its surroundings. The cloud is given solid body rotation with angular momentum of \SI{1.85e51}{\gram\square\centi\metre\per\second}, which corresponds to a ratio of rotational energy to gravitational energy of $\beta_{\mathrm{rot}} \approx 0.01$ \citep{lewis18}. It is important to note that the initial conditions are identical across the three models, with the sole exception of the magnetic field structure. The construction of the magnetic field in each of the three models is summarised in Table \ref{sims}, and visualised in the top panel of Figure \ref{initcond}.
\subsubsection{Constructing the Uniform Magnetic Field} \label{uni}
In the {\simu} case, the total magnetic field threaded through this cloud is $100\,\mu\mathrm{G}$ in magnitude. This field is oriented parallel to the rotation axis of the cloud and evenly distributed. The mass-to-flux ratio is given by $(M/\Phi)/(M/\Phi)_{\mathrm{crit}} = 5.2$, where the critical mass-to-flux ratio is $\SI{487}{\gram\per\square\centi\meter}\,\mathrm{G^{-1}}$ as defined by \citet{mouschovias76}.
 \begin{table}
   \centering
   \setlength{\tabcolsep}{2pt}
   \begin{tabular}{@{}llll@{} }
      \hline
      \footnotesize
      Simulation  Name  & Uniform & Turbulent & Total RMS \\
       & Component & Component & Field Strength\\
      \hline
     \emph{Uniform} & 100\% & - & $100 \ \mu\mathrm{G}$ \\
      \emph{Partially Turbulent} & 50\% &  50\% & $100 \ \mu\mathrm{G}$\\
      \emph{Fully Turbulent} & -  & 100\% & $100 \ \mu\mathrm{G}$ \\
      \hline
      \normalsize
   \end{tabular}
   \caption{Construction of the magnetic field in each simulation. All other parameters are identical in each model (see Section \ref{init} for details). Columns: (1) Simulation name, (2) Fraction of the field which is uniform, (3) Fraction of the field which is un-ordered (turbulent), (4) Total rms field strength.}
   \label{sims}
\end{table}
\subsubsection{Constructing the Partially Turbulent Magnetic Field} \label{partial}
In the {\simp} case the magnetic field has both an initially uniform and turbulent components. The turbulent component is constructed using the same method as is described in Section \ref{turb}. The overall magnitude of the field remains constant and consistent with the other two cases. The total field is given by
\begin{align}
	B=B_0 + \delta B
\end{align}
where $B_0$ is the uniform component and $\delta B$ is the perturbation or turbulent field component. We set the magnetic energy (proportional $B^2$) of the uniform and turbulent components to be equal. Consistency with our other two cases requires the total field to be $100 \ \mu \mathrm{G}$, so
\begin{align}
	\langle B^2\rangle=(100 \ \mu \mathrm{G})^2
\end{align}
which gives
\begin{align}
	B_0=&\frac{\sqrt{2}}{2}\cdot100 \ \mu \mathrm{G}\\
	\left[\langle(\delta B)^2 \rangle\right]^{\frac{1}{2}}=&\frac{\sqrt{2}}{2}\cdot100 \mu \mathrm{G}
\end{align}
giving each component a value of $\sim70.7 \ \mu \mathrm{G}$.

\subsubsection{Constructing the Fully Turbulent Magnetic Field} \label{turb}
As outlined in Section \ref{init}, the {\simp} and {\simt} cases have turbulent magnetic field structures. We do not \emph{drive} turbulence in the field, but rather construct the initial conditions such that the field vectors are randomly orientated. Here we employ a Kazantsev power-law spectrum with an exponent of $3/2$ as the Fourier decomposition of the turbulent field \citep{brandenburg05, federrath16}. The basic Fourier technique for constructing turbulent vector fields follows the method described in \citet{federrath10a}. The range of the wavenumber $k$, in units of $2 \pi /L$, is within $2<k<10$, where $L$ is the length of the box. This constrains the spectrum to the range $4\pi / L\text{ to } 20\pi / L$. The Kazantsev spectrum is the result of turbulent dynamo amplification \citep{kazantsev68,federrath11b}. This is particularly important because field amplification via the turbulent dynamo acts initially on the small scale seeds of a magnetic field \citep{brandenburg05,schober12a,schober12b,schleicher13}.

\section{Results} \label{res}
As the proper time elapsed in each simulation varied between our models, we have chosen to compare the three simulations matched on star formation efficiency rather than time. We define SFE as
\begin{align}
\mathrm{SFE} = \frac{M_{\mathrm{sinks}}}{M_{\mathrm{tot}}},
\end{align}
where $M_{\mathrm{sinks}}$ is the total mass contained in sink particles and $M_{\mathrm{tot}}$ is the initial gas mass of the cloud (1{\msun}). This is discussed further in Section \ref{dis}.
\subsection{Time Evolution of Protostellar Mass}
\begin{figure}
	\centerline{\includegraphics[width=\linewidth]{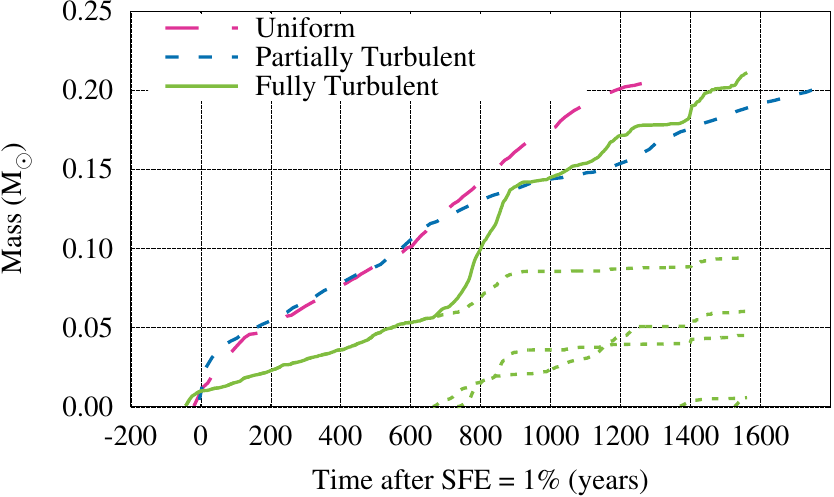}} 
	\caption{Evolution of the sink particle masses in each of the three models. The magenta long-dashed line is the {\simu} case and the blue short-dashed line is the {\simp} case. The solid green line is the {\simt} model and shows the sum of accreted mass in each sink particle, while the dotted lines indicate the mass evolution of each individual sink particle in this model. Each model is scaled such that $\mathrm{time}=0$ corresponds to the time when $\mathrm{SFE} = 1\%$ in each model. As shown in the figure, each simulation was run until the mass in sink particles reached 0.2{\msun} or $\mathrm{SFE} = 20\%$.}
	\label{sink_evol}
\end{figure}

As the simulations evolve, the spherical gas cloud collapses into an accretion disc in the centre of the domain. The disc structure is a fundamental result of the conservation of angular momentum. Within this protostellar disc, sink particles form. The {\simu} case formed exactly one sink particle at the centre of the domain. The {\simp} case also produces a singular sink particle, approximately $8\,$AU off centre. The {\simt} case produces a total of five sinks, each forming at different times (Figure \ref{sink_evol}). The mass evolution of the sink particles in each model is shown in Figure \ref{sink_evol}.

The first and most massive sink particle in the {\simt} case forms $9\,$AU from the centre, with each subsequent particle forming within $52\,$AU of the first. Perturbations in the gas caused by the turbulent magnetic field result in real fragmentation of the disc, evident in the multiplicity of the {\simt} model and the off-centre sink formation in the {\simp} case.

The turbulent magnetic field produces a more isotropic magnetic pressure across the domain. This has implications for the magnitude of the magnetic pressure gradient at the epicentre of collapse as the gas is less efficient at condensing the field lines in the disc, delaying star formation in the {\simt} case. Consequently, there is initially low SFE, which is illustrated by the shallow gradient of the {\simt} model in Figure \ref{sink_evol}. SFE reaches 20\% after $\sim\!1200$, $\sim\!1800$, $\sim\!1500$ years in the {\simu}, {\simp} and {\simt} cases, respectively. 

In summary, we find that the turbulent field component has a substantial impact on the evolution of the protostellar mass and on the fragmentation of the disc.

\subsection{Jet and Outflow Morphology}
\begin{figure*}
	\centerline{\includegraphics[width=\linewidth]{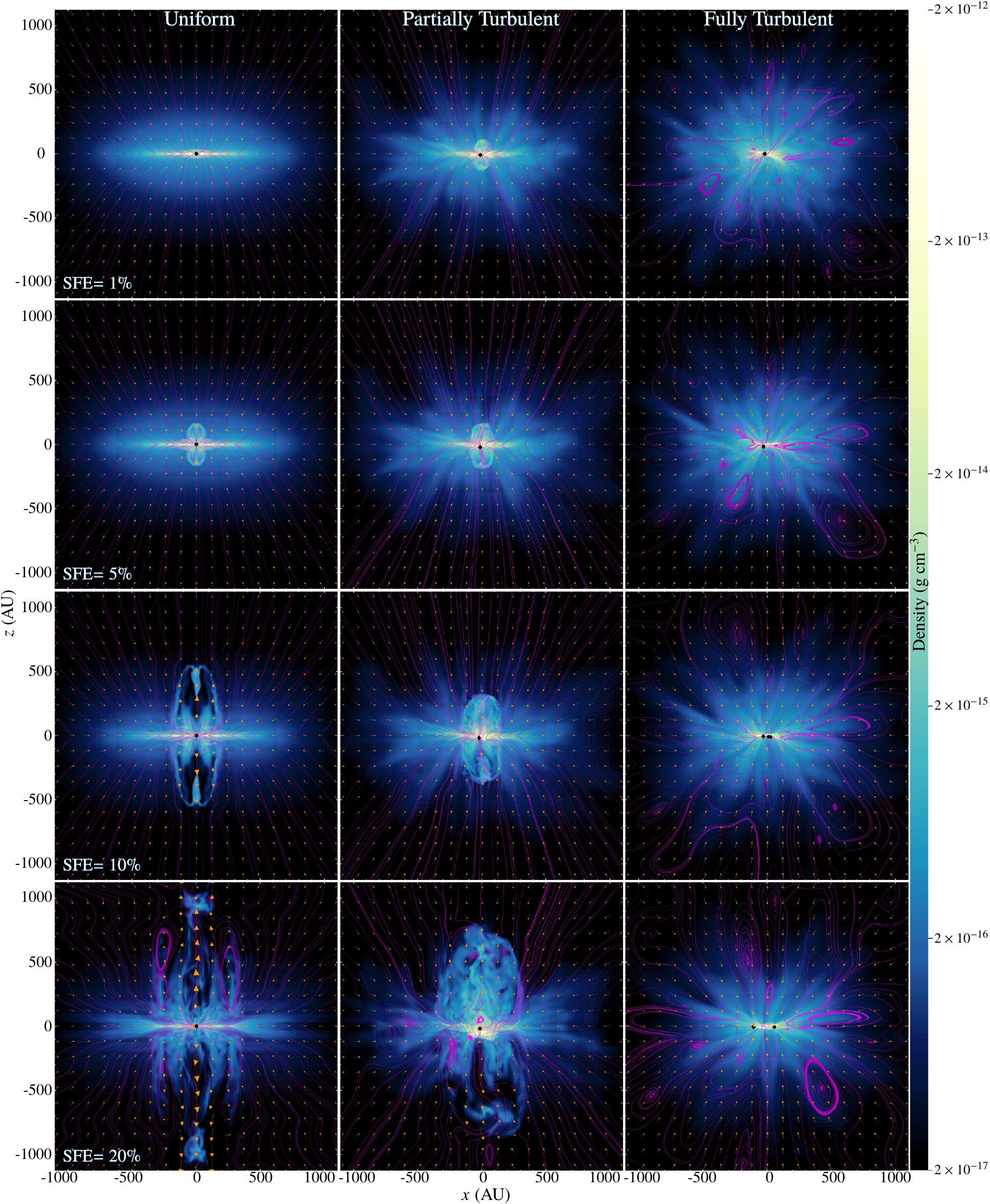}}
	\caption{Visualisations in the \emph{xz}-plane of the evolution of each model. They are matched on SFE, such that `time' increases from top to bottom.  The colour scale shows the average gas density within a three-dimensional slice of dimensions (2200,100,2200)\,AU. Each slice is centred on (0,0,0)\,AU. The left column shows the evolution of the {\simu} model, the middle column shows the {\simp} model and the right column shows the {\simt} model. The magenta streamlines trace the magnetic field structure. The orange vectors represent the velocity field. The scaling of the vectors goes as the square root of the magnitude of the velocity, such that an arrow of length $100\,$AU corresponds to a velocity of \SI{10}{\kilo\meter\per\second}. Sink particles are shown as black crossed spheres.}
	 \label{evolution}
\end{figure*}

In Figure \ref{evolution} we visualise the evolution of density slices, with magnetic field lines and velocity vectors superimposed. The colour scale in Figure \ref{evolution} shows the average density in the \emph{xz}-plane of a slice of the domain. The slices are centred on (0,0,0), and have dimension of $\pm1100 \,$AU in the \emph{x}-, $\pm100 \,$AU in the \emph{y}- and $\pm1100 \,$AU in the \emph{z}-direction.

Figure \ref{evolution} illustrates that the {\simu} case and the {\simp} case produce definite jets and outflows, respectively. As expected, the {\simu} case produces collimated jets consistent with previous works, whereas collimation does not dominate the outflows of the {\simp} case, which are asymmetric and `lobe-like'. Most importantly, we find that jets are completely absent in the {\simt} case. The mechanisms that drive the variance in jet and outflow morphology are discussed in Section \ref{launch}.

\subsection{Jet and Outflow Efficiency}
\begin{figure}
	\centerline{\includegraphics[width=\linewidth]{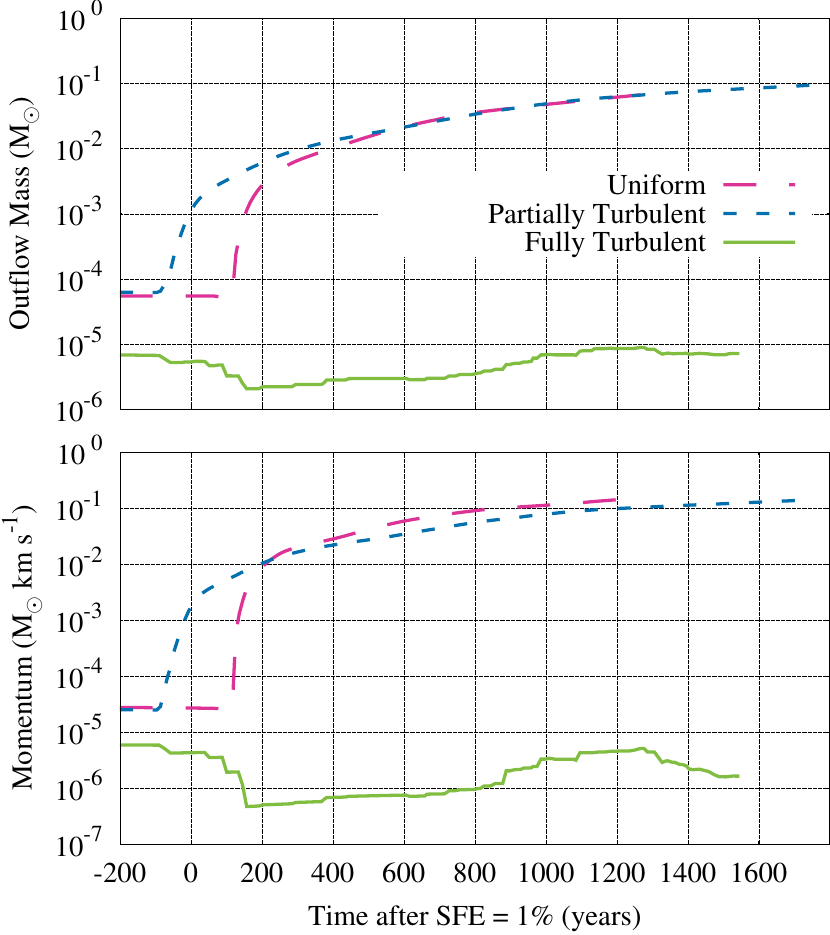}}
	\caption{The time evolution of outflowing mass (top) and outflow momentum (bottom) for each model, beginning when $\mathrm{SFE} = 1\%$. Jets in the {\simu} case (magenta long-dashed line) launch around 100 years after $\mathrm{SFE} = 1\%$. Consistent with Figure \ref{evolution}, outflows in the {\simp} case (blue short-dashed line) begin to form 100 years before $\mathrm{SFE} = 1\%$. The {\simp} case is less efficient in momentum but similar in outflowing mass to the {\simu} case, and there is no distinct outflowing material in the {\simt} case (green solid line).}
	\label{outflow}
\end{figure}

Figure \ref{outflow} corroborates the visual evidence in Figure \ref{evolution}. Here we see the behaviour of material in two cylindrical analysis volumes defined as follows. The two cylindrical volumes begin $100\,$AU above and below the centre of mass of the system, with a radius of $500\,$AU and height of $1000\,$AU. As Figure \ref{evolution} shows, these two cylinders well encompass the outflow region, enabling us to quantify the mass and momentum of the outflowing material. Outflowing material is defined as any cell in which $v_z > 0 \text{ for } z>0$ and $v_z < 0 \text{ for } z<0$. 
As indicated in the top panel of Figure \ref{outflow}, jets begin around 100 years after SFE reaches $1\%$ in the {\simu} case, while in the {\simp} case the outflow begins 100 years before. Although the {\simp} outflows start first, the amount of outflowing mass in the {\simu} case quickly matches that in the {\simp} case. The momentum of the {\simu} jets similarly exceeds the {\simp} outflows after 300 years. This confirms the visual evidence in Figure \ref{evolution} that the {\simp} case is less efficient at expelling mass from the disc than the {\simu} case. 500 years after $\mathrm{SFE} = 1\%$ the outflowing material has characteristic velocities of \SI{3.16}{\km\per\second} in the {\simu} case, \SI{1.78}{\km\per\second} in the {\simp} case and  \SI{0.32}{\km\per\second} in the {\simt} case. These outflow speeds are lower than the $>10\,\mathrm{km}\,\mathrm{s}^{-1}$ often found in observations, however, the outflows here are measured relatively close to the disc and at early times; they are still gaining speed and momentum. The outflow speeds in the {\simu} and {\simp} cases are consistent with the speeds measured in other numerical studies of first core outflows \citep{tomida10,bate14,wurster18a,vaytet18}. For example, in comparable simulations \citet{bate14} find outflow speeds from the first core to be $1.2\,$-\SI{1.8}{\km\per\second} and $1.0\,$-\SI{2.5}{\km\per\second} for initial field strengths of $81\mu\mathrm{G}$ and $163\mu\mathrm{G}$, respectively.

To compare the outflow efficiency of the {\simp} and {\simt} models to that of the {\simu} model, we take the ratio of the time averaged value of the outflow quantities for each pair. That is,
\begin{align}
\langle q(t) \rangle = \frac{\int_{t=-200\,yr}^{t=1200\,yr} q(t) dt}{\int_{t=-200\,yr}^{t=1200\,yr} dt}
\end{align}
where $q(t)$ is outflow mass or momentum. As the {\simu} case is intended to be the control run, we quantify the efficiency of each model by comparing $\langle q(t) \rangle$ for mass and momentum in the {\simp} and {\simt} cases to the {\simu} case. We find that the relative fraction of outflowing mass in the {\simp} and {\simt} cases are $\sim\!100\%$ and $\sim\!0.022\%$, respectively. Compared to the {\simu} model, the momentum of outflows is $\sim\!71\%$ and $\sim\!0.0045\%$ for the {\simp} and {\simt} cases. This is summarised in Table \ref{eff}. While the amount of outflowing mass in {\simu} and {\simp} cases is comparable, the {\simp} case takes approximately 200 years longer to achieve this, making the jet in the {\simu} case the most efficient at transporting mass away from the disc. This is due to the high speeds of the jet material in the {\simu} case. In contrast, we find no significant outflow in the {\simt} case, making it the least efficient at ejecting material.
\begin{table*}
   \centering
   \begin{tabular}{@{} l l l l l @{} }
      \hline
      \footnotesize
      Model  \ \ & Outflow Mass Efficiency \ \ & Outflow Momentum Efficiency & Sink Formation Time & Outflow Launching Time \\
      \hline
     \emph{Uniform} & 1.00 & 1.00 & 35120  years & 35260  years\\
      \emph{Partially Turbulent} & 1.00 & 0.71& 35353 years & 35273  years\\
      \emph{Fully Turbulent} & \SI{2.15e-4}{} & \SI{4.45e-5}{} & 34943 years & - \\
      \hline
      \normalsize
   \end{tabular}
   \caption{Normalised efficiency of outflow quantities. The quantities in the {\simu} model are normalised to $100\%$ efficiency so that the relative efficiency of the {\simp} and {\simt} models can be calculated. Columns: (1) Model name, (2) Relative fraction of outflow mass, (3) Relative fraction of outflow momentum, (4) Sink formation time (absolute time), (5) Launching time (absolute time).}
   \label{eff}
\end{table*}

\section{Discussion} \label{dis}
Our study is designed to isolate and examine the contribution of initial magnetic field structure, to the complex balance of forces that govern star formation. As shown in Section \ref{res}, only the models with some ordered, poloidal component of the field produce jets and outflows.
\subsection{Jet Launching Mechanisms} \label{launch}
To explain the variance in outflow efficiency and morphology between our three models, we must examine the physical mechanisms by which protostellar jets and outflows are launched. It is possible to launch outflows via the force provided by a magnetic pressure gradient \citep{lynden-bell03}, although the morphology of an outflow driven solely by this mechanism need not be collimated or orientated perpendicular to the accretion disc. Collimated jets are driven by the \emph{magneto-centrifugal mechanism} \citep{blandford82}. In this model, the rotation of the accretion disc `coils up' the magnetic field, concentrating the field lines in the disc which provides a magnetic pressure gradient, as well as a centrifugal force from the rotation. This mechanism is described by \citet{blandford82} to form jets only if the poloidal component of the field makes an angle of $\ge 30^{\circ}$ with the rotation axis, and the poloidal component of the field above and below the disc dominates any toroidal contribution to the field within the disc. The combination of these forces accelerates the gas along the field lines and forms collimated bipolar jets along the rotation axis of the disc. 
\begin{figure*}
	\centerline{\includegraphics[width=\linewidth]{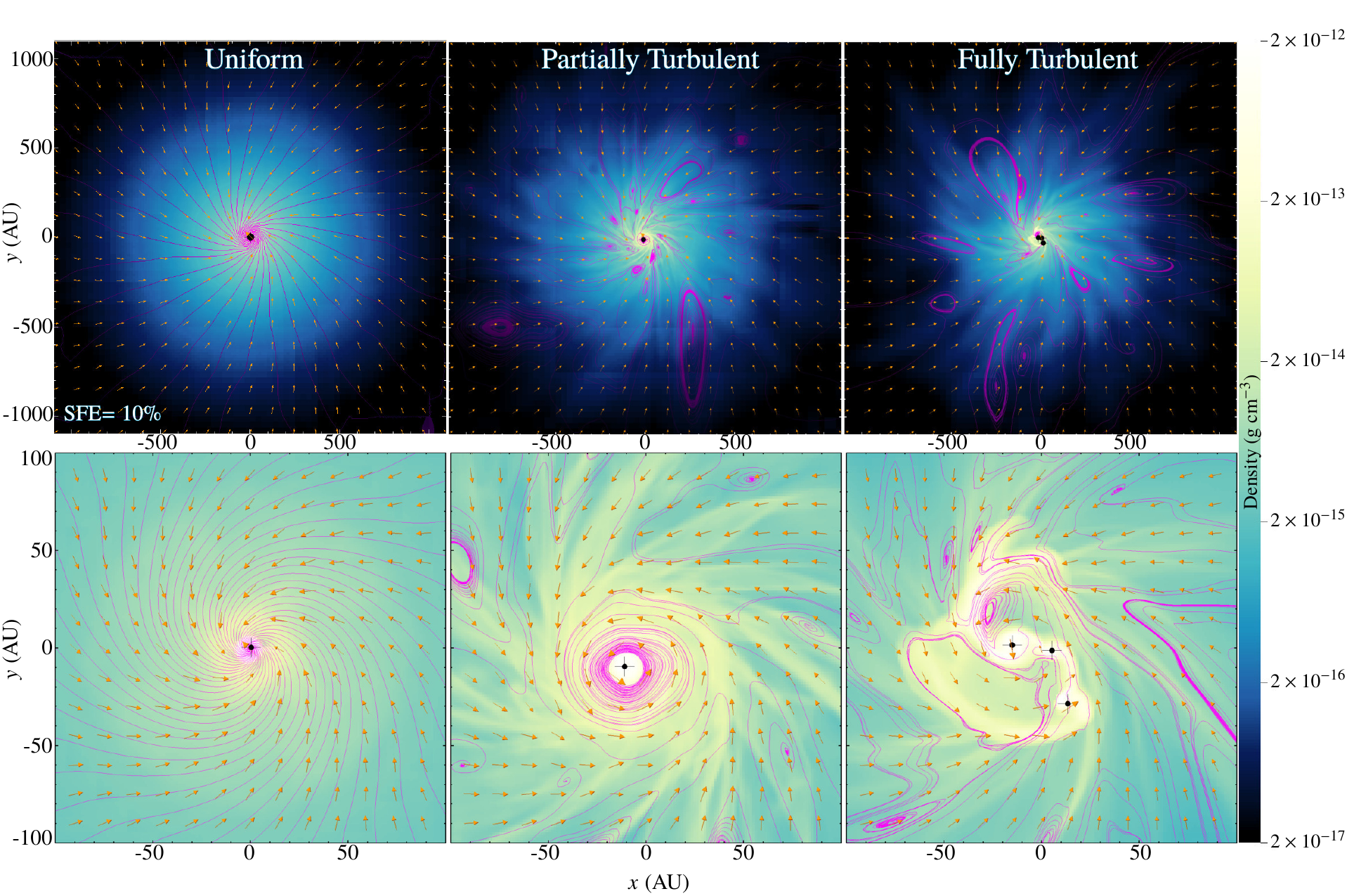}}
	\caption{Similar to Figure \ref{evolution} but in the \emph{xy}-plane. The characteristic time of  $\mathrm{SFE} = 10\%$ is shown for each simulation. In both of these panels, the slice in the \emph{xy}-plane (with thickness of $\Delta z = 100\,\mathrm{AU}$ in all cases) is centred on the \emph{z}-position of the sink particle (or on the average \emph{z}-position of the three sink particles in the {\simt} case). The slice is centred around (0,0) in the \emph{x} and \emph{y} directions. The colour bar shows average gas density in the slice. \emph{Top}: As in Figure \ref{evolution}, th dimensions of these panes are  (2200,2200,100)\,AU. \emph{Bottom}: A zoomed-in view of the top panel, with dimensions (200,200,100)\,AU. It should be noted that the structure of the magenta streamlines representing the magnetic field lines depends on the integration, domain, sampling rate and step size chosen. We have chosen these conditions in such a way as to best represent the underlying physical structure of the field.}
	 \label{xyplane}
\end{figure*}

Given the efficiency and high degree of collimation of jets in the {\simu} model, evident in Figure \ref{evolution}, and given that the work of \citet{blandford82} assumed a uniform field oriented along the rotation axis of the disc, we conclude that this is the driving mechanism of the jets in our {\simu} model. This is corroborated by the tightly coiled structure of the field lines, seen in Figure \ref{xyplane} which shows the same features as Figure \ref{evolution} at $\mathrm{SFE} = 10\%$ but in the \emph{xy}-plane. Our {\simp} model produced asymmetric, wide-angle outflows which were less efficient at transporting mass and momentum away from the disc than the jets in the {\simu} model. We conclude that the launching mechanism of the outflows in the {\simp} case is dominated by magnetic pressure gradients, although a weak magneto-centrifugal mechanism is likely still contributing since there is some poloidal component to the field, and the outflows are weakly collimated. In this case, the field lines are less uniformly and tightly coiled (Figure \ref{xyplane}), which suggests that the magneto-centrifugal mechanism plays a lesser role in launching the outflow than in the {\simu} case.

As seen in Figure \ref{evolution} and \ref{outflow}, there was no distinctly outflowing material in the fully turbulent model. At $\mathrm{SFE} = 20\%$, it is evident in the velocity field that material is still predominantly infalling. We conclude that the chaotic field structure has inhibited the launching of any outflow in the {\simt} case. The absence of a uniform poloidal field component, not the presence of the turbulent field, inhibits the \citet{blandford82} mechanism and does not allow for the formation of fast collimated jets. This is corroborated by the lack of a coiled field structure in Figure \ref{xyplane}. However, we also did not find any significant less collimated outflow in this case; this indicates that the magnetic pressure gradient is not (yet) strong enough to revert the collapse. This implies that a poloidal field component is essential for jet and outflow formation. 

\begin{figure*}
	\centerline{\includegraphics[width=\linewidth]{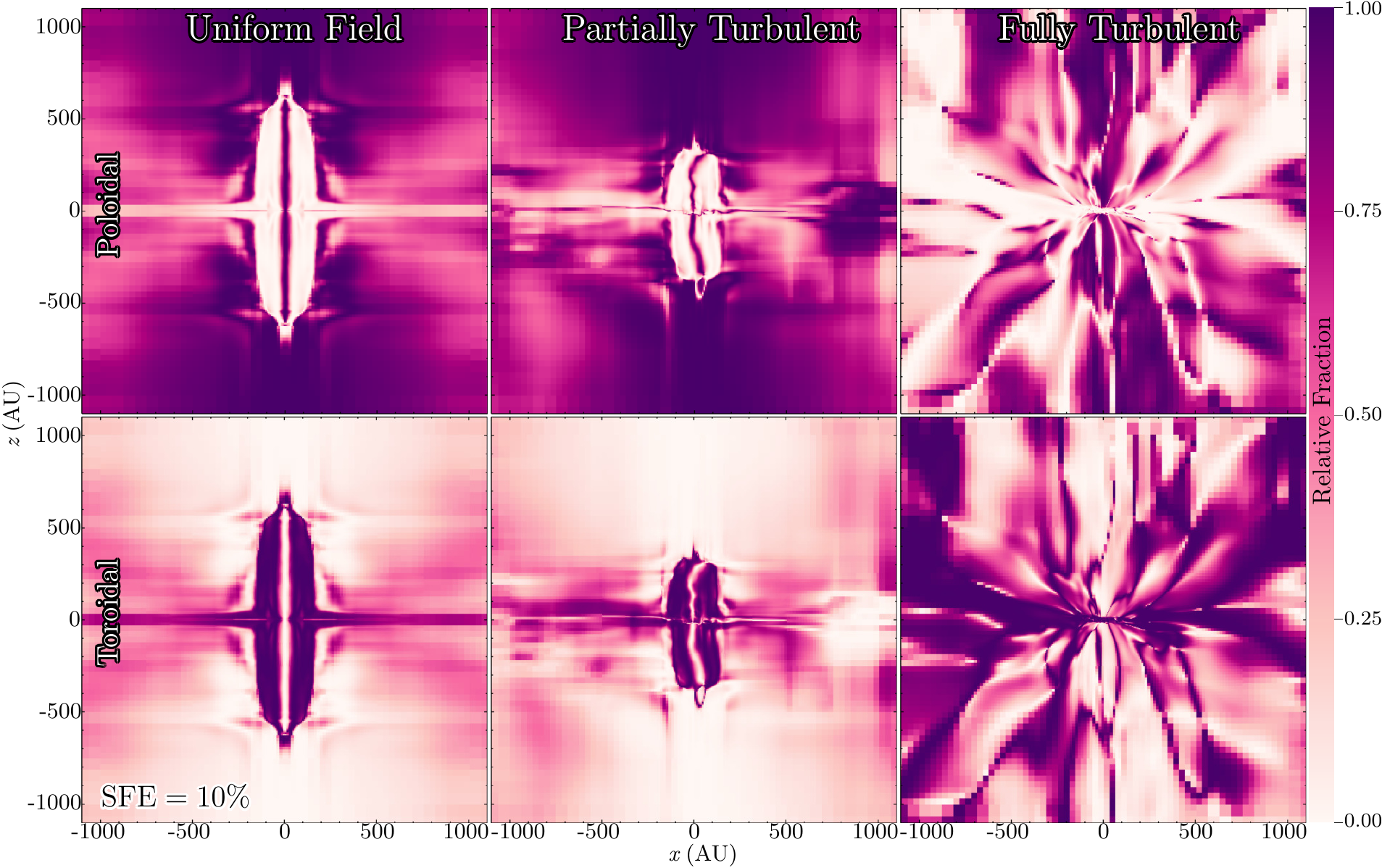}}
	\caption{The poloidal and toroidal field components as a relative fraction of total field strength, shown in the \emph{xz}-plane. These values are the average field strength in a slice of the same size as in Figure \ref{evolution}. \emph{Top}: The poloidal field fraction for each case, $B_z^2/(B_x^2+B_y^2+B_z^2)$. \emph{Bottom}: The toroidal field fraction for each case, $(B_x^2+B_y^2)/(B_x^2+B_y^2+B_z^2)$.}
	\label{polvtol}
\end{figure*}

Figure \ref{polvtol} shows the relative fraction of poloidal and toroidal field components to the total field strength in each of the three simulations. It becomes evident that the {\simt} case contains no clearly ordered poloidal or toroidal structure in its field. Conversely, Figure \ref{polvtol} reveals the poloidal field at the centre of the jet in the {\simu} model and also in the outflow in the {\simp} model. The prominent toroidal, `cage-like' structure expected as a result of the winding of field lines in the magneto-centrifugal mechanism is also seen in these two cases. Figure \ref{xyplane} and Figure \ref{polvtol} combined make a convincing argument for the jets and outflows in the {\simu} and {\simp} models being driven by a combination of magnetic pressure gradients and the magneto-centrifugal mechanism.

\subsection{Star Formation and Multiplicity} \label{multi}
In the {\simt} case, the turbulent magnetic field causes perturbations in the gas density due to locally varying magnetic pressure gradients. This perturbs the cloud, but an additional consequence is the isotropy of magnetic pressure, as compared to the {\simu} and {\simp} cases, because the field has no mean component. In the context of the work of \citet{price07}, fragmentation is an expected consequence of this density perturbation, but the time-delay of multiple protostar formation and accretion is also expected, due to the extra support of the isotropic magnetic pressure. This core ultimately fragments into five sink particles (as seen in Figure \ref{sink_evol}). The sink particle formulation discussed in Section \ref{sinks} ensures that this fragmentation is not a numerical artefact. Combined with the large body of evidence that multiplicity is ubiquitous in star formation \citep{duchene13,bate15}, it is unsurprising that the {\simt} model produced multiple protostars.

Although we have chosen to match our simulations on SFE rather than absolute time, we can report that there is a significant time delay of 230 years between the formation of the sink particles in the {\simu} and {\simp} cases, with the protostar forming later in the {\simp} case.  In the reference time frame of Figure \ref{outflow}, the relative time difference between jet and outflow launching in these two simulations is also of the order of 200 years. This means that the outflows in each case form at approximately the same absolute time, and that the sink particle in the {\simp} case forms after its outflow is launched. However, SFE quickly reaches 1\%, explained by the strong initial phase of accretion as evident in Figure \ref{sink_evol}. The delay in sink formation and the high accretion rate in the {\simp} case are a result of the turbulent component of the magnetic field. As in the {\simt} case, the isotropy of magnetic pressure supports the core against collapse, which allows material to be concentrated in the disc while suppressing star formation. Eventually, the gravitational force overcomes the magnetic pressure, and a sink particle forms. There is then a large reservoir of material available for accretion, resulting in a steep increase in SFE. 

\subsection{Caveats} \label{cav}
\subsubsection{Non-Ideal MHD}\label{nonideal}
Our calculations do not include non-ideal MHD effects, which can lead to the suppression or distortions of outflows \citep[e.g.][]{ wurster16,wurster18b}. The smallest scale resolved in our simulations is $\sim\!3\,$AU, which is larger than the scales on which Ohmic diffusivity is important \citep{shu06,konigl11}. The Hall effect dominates on scales of $1\!-\!10\,$AU \citep{konigl11}, but its influence on outflow formation is strongly dependant on the angle between the magnetic field and rotation axis of the cloud (e.g. see \citet{wardle99} for an analytic discussion; see numerical examples in \citet{tsukamoto15,wurster16,wurster18c}). The random orientation of the initial magnetic field in the {\simt} case could result in regions of the field being anti-aligned with the rotation axis of the cloud. In such a scenario the inclusion of Hall effect may cause the outflow of material to be suppressed \citep[e.g.][]{wurster18c}, but we do not expect that this effect would increase the likelihood of outflows developing in the {\simt} case. 

The effects of Ohmic resistivity and ambipolar diffusion are independent of the orientation of the magnetic field. For this reason we expect that our results would be uniformly impacted by the inclusion of these effects, regardless of the initial field morphology. Simulations by \cite{wurster16} show that the inclusion of Ohmic resistivity has little impact on the formation jets or their structure. It should be noted that ambipolar diffusion, which dominates on larger scales where the density is low and the fractional ionisation is high \citep{braiding12}, may suppress outflows and effect their morphology, although they are still expected to form at later times \citep{vaytet18}. Non-ideal MHD calculations are computationally intensive \citep{nolan17,wurster18b}, so we leave it to follow-up studies to include non-ideal MHD and discuss the results in the context of this work. However, the fact that all our simulations use the same ideal MHD approximation still gives a qualitative sense of the relative importance of the magnetic field structure on disc morphology and jet/outflow properties.
\subsubsection{The Effects of Radiation}
Radiation is an important aspect of star formation. Radiative feedback has been found to hinder disc fragmentation in some cases \citep{offner09,price09,kuiper16,federrath17,guszejnov18}, while radiation transport plays a role in shaping the temperature and density structure of the protostellar disc. The inclusion of radiative effects in our study may have changed the absolute structure of the discs in each of our models. It is not, however, expected to emphasise the importance of the magnetic field structure in any of our models (relative to each other), when compared to simulations without radiative effects. 
\subsubsection{Numerical Resolution} \label{numres}
The highest level of refinement used in our simulations is $L_{\mathrm{ref}}=11$ which corresponds to a cell length of $3.05\,$AU on the highest level of AMR. The minimum refinement is $L_{\mathrm{ref}}=4$, the length of these cells being $125\,$AU. We chose this level of refinement based on convergence tests carried out by \citet{kuruwita17}, given that our initial setup closely resembles that work. \citet{federrath14} find that full convergence in a domain of this size would require $L_{\mathrm{ref}}=17$. This would result in sub-AU resolution ($\Delta x = 0.6\,$AU), and enable us to better resolve the launching velocities of the jets/outflows. Although the absolute outflow mass and momentum are not converged at $L_{\mathrm{ref}}=11$, the outflow quantities in the {\simp} and {\simt} cases as a fraction of the {\simu} case are nonetheless representative of the same relative fractions that would be computed at higher resolutions. Future studies should aim to investigate the results of this work in simulations with higher levels of refinement. Increasing the resolution would also benefit the investigation of non-ideal effects, as discuss in Section \ref{nonideal}.

\section{Conclusions}\label{conc}
We have conducted a systematic numerical study of the effects of uniform and turbulent magnetic fields on the formation of bipolar jets and outflows from protostellar discs, and analysed their morphology and efficiency. The behaviour of each of the three cases is as follows:
\begin{enumerate}
	\item The {\simu} case produces a highly collimated jet and forms a single protostar.
    The jet is launched via a magneto-centrifugal mechanism at a similar time to protostar formation. 500 years after $\mathrm{SFE} = 1\%$ the outflowing material has characteristic velocities of \SI{3.16}{\km\per\second}. The jet in this model is the most effective at transporting mass and momentum away from the disc, causing the initial accretion phase to be less efficient than in the {\simp} case. 
        
    \item The {\simp} case produced definite outflows with a weakly collimated, asymmetric morphology. The outflow was launched before SFE reached 1\%, likely due to strong initial accretion. 500 years after $\mathrm{SFE} = 1\%$ the outflowing material has characteristic velocities of \SI{1.78}{\km\per\second}. The launching mechanism in this case is likely a combination of weak magneto-centrifugal forces and magnetic pressure gradient. The wide angle outflows in this model were comparable to that of the {\simu} case in mass, and 29\% less efficient in momentum. This model formed a single protostar. 
    
    \item The {\simt} case produced no definite bipolar outflows, and the fractional transport efficiency in comparison to the {\simu} case is negligible (0.021\% in mass and 0.0045\% in momentum). 500 years after $\mathrm{SFE} = 1\%$ the outflowing material has characteristic velocities of \SI{0.32}{\km\per\second}. The turbulent magnetic field induced fragmentation of the disc in this model, forming five protostars in total. Accretion was initially inhibited by the isotropy of magnetic pressure. 
\end{enumerate} 

This study explored the extrema of the effects of magnetic field structure on jet and outflow formation (fully ordered to fully turbulent). The {\simp} case is the most physically realistic, and future work should be done to examine the parameter space around this midpoint, including a partially turbulent field with a misaligned uniform, poloidal component of the kind discussed in Section \ref{intro}.
\section*{Acknowledgements}
We thank the anonymous referee for their valuable suggestions which helped us to improve this paper. IG would like to thank the Australian National University and the Research School and Astronomy and Astrophysics for the opportunity and funding provided as part of the Summer Research Scholarship program. RK thanks the Australian government for the Australian government research training program scholarship stipend. CF acknowledges funding provided by the Australian Research Council's Discovery Projects (grants ~DP150104329 and ~DP170100603), the Future Fellowship scheme (grant FT180100495), and the Australia-Germany Joint Research Cooperation Scheme (UA-DAAD). The simulations and data analyses presented in this work used high-performance computing resources provided by the Leibniz Rechenzentrum and the Gauss Centre for Supercomputing (grants ~pr32lo, pr48pi and GCS Large-scale project ~10391) and from the Australian National Computational Infrastructure (grant ~ek9, fu7) in the framework of the National Computational Merit Allocation Scheme and the ANU Allocation Scheme. The simulation software FLASH was in part developed by the DOE-supported Flash Centre for Computational Science at the University of Chicago. Visualisations were made using Visit \citep{visit}. Analysis was performed using yt \citep{turk11}.

%\bibliographystyle{mnras}
%\bibliography{bella} 

% Don't change these lines
\bsp	% typesetting comment
\label{lastpage}
\end{document}